# CHANNELS REALLOCATION IN COGNITIVE RADIO NETWORKS BASED ON DNA SEQUENCE ALIGNMENT

Santosh Kumar Singh[1], Dr. Krishna Chandra Roy[2] and Vibhar Pathak[3]

[1]Department of Computer Engineering, Suresh Gyan Vihar University, Jaipur, India
sksmtech@yahoo.com
[2]Department of ECE, SBCET, Benad Raod, Jaipur, India
roy.krishna@rediffmail.com
[2]Department of Information Technology, Suresh Gyan Vihar University, Jaipur, India
vibhakar_p@yahoo.com

## ABSTRACT

*Nowadays, It has been shown that spectrum scarcity increased due to tremendous growth of new players in wireless base system by the evolution of the radio communication. Resent survey found that there are many areas of the radio spectrum that are occupied by authorized user/primary user (PU), which are not fully utilized. Cognitive radios (CR) prove to next generation wireless communication system that proposed as a way to reuse this under-utilised spectrum in an opportunistic and non-interfering basis. A CR is a self-directed entity in a wireless communications environment that senses its environment, tracks changes, and reacts upon its findings and frequently exchanges information with the networks for secondary user (SU). However, CR facing collision problem with tracks changes i.e. reallocating of other empty channels for SU while PU arrives. In this paper, channels reallocation technique based on DNA sequence alignment algorithm for CR networks has been proposed.*

## 1. INTRODUCTION

The distribution of radio spectrum is under control of the central government; the Federal Communications Commission (FCC) published a report in November 2002, prepared by the Spectrum-Policy Task Force, aimed at improving this precious resource. The allocation of the unlicensed frequency bands has resulted in the overcapacity of these bands. The most of the usable frequency spectrum already has been assigned for licensed user, resulting in a scarcity of spectrum for new and up-and-coming wireless applications. To resolve this crisis, regulators and policy makers are working on new spectrum management strategies. Particularly, the U.S. Federal Communications Commission (FCC) is tackling the problem in three ways [1]: spectrum reallocation, spectrum leases, and spectrum sharing. In spectrum reallocation, bandwidth from government and other long-standing users is reassigned to new wireless base services such as mobile communication, broadband Internet and video distribution. In spectrum leases, the FCC relaxes the technical and business limitations on existing spectrum licenses by permitting existing licensees to use their spectrum flexibly for various services or even lease their spectrum to third parties. Spectrum sharing has allocation of an unmatched amount of spectrum that could be used for unlicensed or shared service. Where as spectrum reallocation and spectrum leases focused on improving the efficiency of spectrum usage. The FCC is considering a new spectrum-sharing pattern, where licensed bands are opened to unlicensed operations on a non-interference basis. Because some licensed bands (such as TV bands) are under-utilized, spectrum sharing in empty sections of these licensed bands can fill the spectrum shortage problem. In this spectrum-sharing model — which is frequently referred to as dynamic spectrum access (DSA) - licensed users are referred to primary users (PU), whereas unlicensed users that access spectrum opportunistically are referred as secondary users (SU). A snapshot of spectrum utilization up to 6 GHz in urban area of the radio spectrum in the revenue-rich urban areas as shown in figure 1, it is found that 1) Some frequency bands in the spectrum are mostly vacant most of the time; 2) Some other frequency bands are only partially busy;

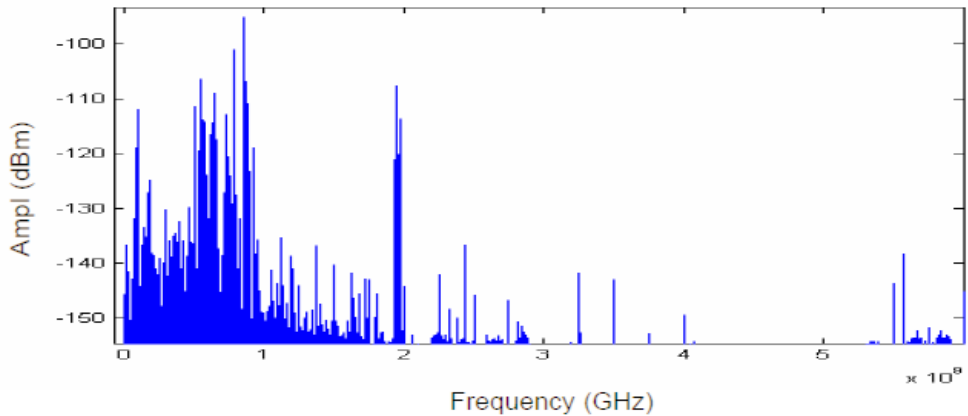

Figure 1

3) The remaining frequency bands are heavily used. The under utilization of the electromagnetic spectrum leads to think in terms of spectrum holes, which is defined by the author in [2]: A spectrum hole is a band of frequencies assigned to a primary user, but, at a particular time and particular geographic location, that user is not utilizing the band. The spectrum utilization can be improved notably by making it possible for a secondary user to access a spectrum hole that unoccupied by the primary user. The spectrum holes have been utilized to promote the efficient use of the spectrum by take advantage of the existence of spectrum holes. This concept is known as Cognitive radio [3]. Joe Mitola, after introducing the note of software radio in 1991, together with Gerald Maguire, used the term cognitive radio (CR) for the first time in 1999. Mitola and Maguire identify a CR as an improvement of a software radio and state: "Radio protocol is the set of RF bands, air interfaces, protocols, and spatial and sequential patterns that moderate the use of radio spectrum. Cognitive radio extends the software radio with radio-domain model-based reasoning about such protocols." Mitola introduces radio cognition cycle, explicitly elaborated in [3]. A suitable description is found in paper of Haykin [4]: "Cognitive radio is an intelligent wireless communication system that is aware of its surrounding environment i.e. its outside world, and uses the methodology of understanding-by building to learn from the environment and adapt its internal states to statistical variations in the incoming radio frequency (RF) stimuli by making corresponding changes in certain operating parameters e.g. transmit power, carrier frequency, and modulation strategy in real time. A CR is a Software defined radio (SDR) that additionally senses its environment, tracks changes, and possibly reacts upon its findings. The technology of cognitive radio (CR) plays an important role in realizing the DSA paradigm. In addition, a CR can learn from its environment and intelligently adjust its operating parameters based on what was learned. In DSA, CRs used by SU must be able to scan a certain spectrum range and intelligently decide which spectrum band to use for its transmission. During spectrum sensing, if a SU detects that it is within a PU's protection area of a particular band, it stops from accessing that band and searches for an empty band that is accessible called channel reallocation. If no PUs is detected, the SU coordinates with other SU to share the spectrum not utilized by PUs. Depending on the operation scenario, the SU can employ either cellular network architecture or an ad_hoc network architecture. A cellular CR network architecture is employed in the IEEE 802.22 standard that specifies the air interface (physical [PHY] and medium access control [MAC] layers) for a CR-based wireless regional area network (WRAN). A WRAN cell is composed of a base station and a number of consumer premise equipments, and the coverage of a WRAN cell can range from tens of kilometres to a hundred kilometres. In contrast, an ad hoc CR network is comprised of low-energy mobile computing devices equipped with CRs, and they interact with each other via multihop wireless links. The establishment of each wireless link is via DSA. Although these two types of CR networks have different network architectures, spectrum sensing and allocation/reallocation is an essential component of both, and it represents one of the key technological hurdles that must be overcome before the widespread deployment of CR networks is possible.

This paper presents a channels allocation and reallocation optimisation method for secondary user (SU) based on DNA sequence alignment, which is most computationally efficient, compare to other existing complex and crushing technique.

The rest of the paper is organized as follows: Section-2 discusses several other optimisation methods that introduces the problem of spectrum allocation/reallocation and. Section-3 describes DNA sequence alignment optimisation method for channels allocation/reallocation and mat lab simulation results with discussions. Finally conclusion and future work are discussed in Section 4

## 2. PROBLEM FORMULATION

In this section we first review and analyse the drawbacks of general model for spectrum of channels allocation/reallocation optimisation method used in cognitive radio. The objective in of optimization problems is to achieve an optimum solution but the problem is that the search may get complex and one may not recognize where to come across for the solution or where to start with. There are many methods that can help finding a fitting solution but the result might not be the best one. Drawbacks of few of efficient methods are described below. The solutions found using these methods are often considered as good solutions as it is not forever possible to define the optimal.

### 2.1. Game theory

Game theory is an emerging stage and is used for interactive decision conditions provides systematic tools to predict the outcome of composite interactions between the rational entities based on apparent result. It works on predictions of probabilities but stress for a precise awareness of the total number of nodes, but the dynamic nature of real-time networks one doesn't even have the knowledge of what nodes do enter or disappear the network and at real time. Also, the definition of a steady state and fear form an undesirable flow with increasing number of nodes may cause suspicions and may make the realization of Game theory concept in the cognitive radios very complicated. The dimension of the system state in a stochastic game is exponential to the number of users of the system, so that the convergence speed of algorithm suffers badly when the system has large number of cognitive radio users (SU) for channel allocation and reallocation [5].

### 2.2. Fuzzy logic

Another approach is fuzzy logic that offers separate limitations that reason its value in the dynamic nature of the wireless environment, to shrink. It allows the estimated solutions to be establishing in the appearance of uncertain inputs, as cognitive radio networks require certain inputs for channel reallocation to do not interfere (PU). Its logic for approximation does not have an evolutionary ability to allow it to change well in time with the environment encountered at real time [6].

## 2.3. Artificial Neural Networks

Neural networks that are well recognized as an AI technique but it doesn't offer with consistency in terms of guarantee that it will play within a set of operational limitation. Most of the neural networks require for a large training sequence to reproduce the experiential performance and usually perform in unpredicted behavior [6].

## 2.4. Genetic Algorithms

The Genetic Algorithm is easy to apply and can be reutilized to solve additional problems. Implementation of a basic G.A, new object can be added as just another chromosome and using the same encoding scheme just change the existing fitness function and can be solve another optimization problem. However some problems force to find implementation of the encoding scheme for fitness function to be very difficult. The disadvantage of using this algorithm takes much computational time is not mandatory in cognitive radio [6].

Although condition will be worst in above model while SU has already allocated a channel and meantime channel owner PU arrives so that SU have to quit that channel and have to be reallocate another channel [7]. As shown in figure 2 [4], Dynamic spectrum-sharing for four channels, and the way in which the spectrum manager allocates the channel bandwidths for three time instants t1< t2< t3, depending on the availability of spectrum holes. Here once channel has allocated for SU but at next moment PU arrives then reallocation of SU to another band is very complicated task to avoid interference. This reallocation task in the research area of cognitive radio is still challenging. A novel robust and less computational algorithm based on DNA sequence alignment scheme for this task has been proposed in next section.

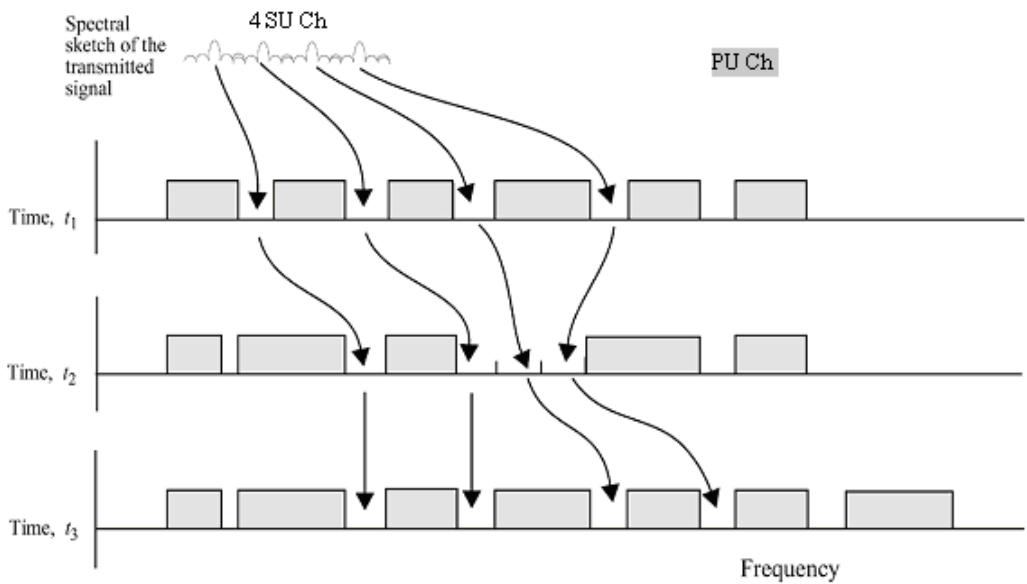

Figure 2

## 3. PROBLEM SOLUTION

In this section we first discuss the background of DNA sequence alignment algorithms and then implementation of proposed solution model on the signal processing of cognitive radio networks along with mat lab simulation results.

### 3.1. Background of DNA sequence alignment scheme

The recent availability of new, less expensive high-throughput DNA sequencing technologies frequently analyses sequence data. These data are being generated for several purposes, including genotyping, genome resequencing, and metagenomics. The standard global sequence alignment algorithm, referred to as Needleman-Wunsch derived from dynamic programming approach [8]. Before discussing this algorithm, it's important to first understand the basic DNA (deoxyribonucleic acid) molecule that are used in computational biology. DNA sequence is preoccupied as a long text over a four-letter alphabet, each representing a different nucleotide: adenosine (A), cytosine (C), guanine (G) and thymine (T). [9]. Needleman-Wunsch (NW) algorithm aligns two DNA sequences by beginning at the ends of the two sequences and attempting to align all possible pairs of characters (one from each sequence) using a scoring scheme for matches, mismatches, and gaps. The highest set of scores defines the optimal alignment between the two sequences [8]. However, the greater the sequence similarity, the greater chance there is that they share similar function and/or structure. Suppose there are two sequences seq1 and seq2 to be aligned, where |Seq1| = m and |Seq2| = n. If gaps are allowed in the sequences, then the potential length of both the first and second sequences is m+n. The similarity between two sequences Seq1 and Seq2 of lengths m and n, respectively, using a NW algorithm approach [10], the initial step is to align one of the sequences across the columns of the matrix, and the other sequence across the rows. The alignment can also start with a gap in one of the sequences, so that care has to be taken as well. Let's assume that the two sequences i.e. Seq1=GAATTCAGTTA and Seq2=GGATCGA have to be aligned, here m=11 and n=7. Since it is possible to begin an alignment with a gap, the algorithm build an (n +1) × (m +1) matrix **S**. Row 0 and column 0 will represent gaps. Rows 1-7 will be labeled with the corresponding nucleotide of the Seq1, while columns 1-11 will be labeled with the corresponding nucleotide of the Seq2. Let **S** be the initial matrix. Now this will need to decide the scoring scheme to be used and requires parameters for a match score, a mismatch score, and a gap score. The match and mismatch scores will be combined into a single match/mismatch score, $s(a_i b_j)$. There will also be a single linear gap penalty score, w. This example has the following parameters:

**Seq1**: GAATTCAGTTA; m = 11, **Seq2**: GGATCGA; n = 7

$s(a_i b_j) = +5$ if $a_i = b_j$ (match score) ; $s(a_i b_j) = -3$ if $a_i \neq b_j$ (mismatch score) ; w = -4 (gap penalty)

Where i = 0, 1, 2, … m and j = 0, 1, 2, ... n

Once the scoring functions set and the sequences to align, there are three steps occupied in calculate the optimal scoring alignment. This methods needs to finish three steps for global sequence alignment as desired. The three steps are as follows:

### 3.1.1. Initialization Step

In the initialization step of global alignment, each row $S_{i,0}$ is set to w * i. In addition, each column $S_{0,j}$ is set to w * j. Here w is the gap penalty and the scoring scheme express above.

### 3.1.2 Matrix Fill Step

The matrix fill step finds the maximum global alignment score by starting in the upper left hand corner in the matrix and finding the maximal score $S_{i,j}$ for each position in the matrix. In order to locate $S_{i,j}$ for any i, j it is least to be familiar with the score for the matrix positions to the left, above and diagonal to i, j. In terms of matrix positions, it is essential to identify $S_{i-1,j}$, $S_{i,j-1}$ and $S_{i-1,j-1}$. For each position, $S_{i,j}$ is defined to be the highest score at position i,j; i.e.

$$S_{i,j} = \text{Max}^m \quad [\ S_{i-1,j-1} + s(a_i,b_j) \text{ (match or mismatch in the diagonal)},$$
$$S_{i,j-1} + w \text{ (gap in Seq1)}, \qquad\qquad (1)$$
$$S_{i-1,j} + w \text{ (gap in Seq2)}\ ]$$

Using equation (1), thus, $S_{1,1}$ = MAX [$S_{0,0}$ + 5, $S_{1,0}$ - 4, $S_{0,1}$ - 4] = MAX [5, -8, -8]. The score at position 1,1 in the matrix can be calculated. Since the first element in both sequences is a G, $s(a_1b_1) = 5$, and by the assumptions declared prior, w = -4. A value of 5 is then placed in position 1,1 of the scoring matrix. Similarly, rest of the matrix can be filled in a similar fashion. The resulting matrix fill is shown in figure 3. Each cell has one to three arrows indicating that which cell the maximum score was obtained.

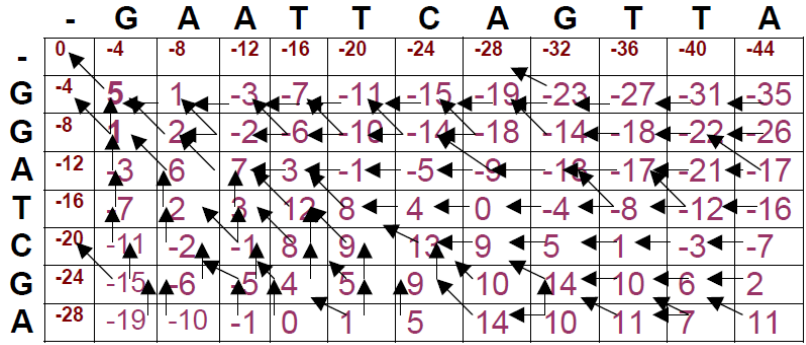

Figure 3

### 3.1.3. Trace-back Step.

After filling the matrix, the maximum global alignment score for the two sequences is 11 at the lower right hand cell value. The trace-back step will obtain the actual alignment that result in the maximum score. The trace-back begins in position $S_{m, n;}$ i.e. the position where both sequences are globally aligned. Since pointers have been kept back to all possible ancestors, the trace-back is simple. At each cell move next according to the pointers. To initiate, the only possible ancestor is the diagonal match as shown in figure 4.

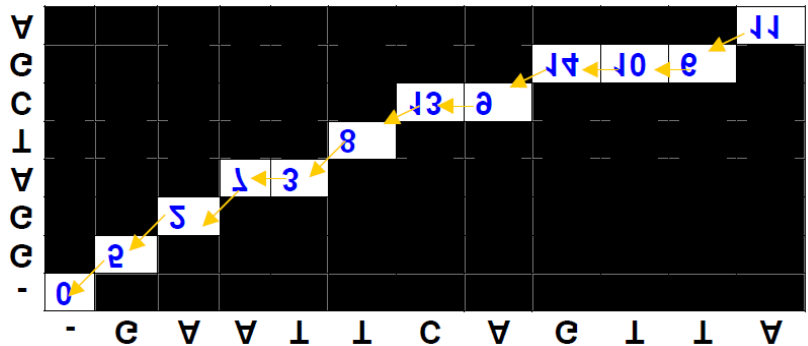

Figure 4

The resultant global alignment of two sequences is as follows:

G A A T T C A G T T A  
|     | |   |     |  
G G A - T C - G - - A

It can easily observe that how the global sequence alignment tool based on Needleman-Wunsch algorithm aligned the two sequences for analysis of their behaviour and characteristics. Therefore this algorithm can be little modified and easily implemented on signal processing of cognitive radio network to set-up efficient relation between PU and SU in terms of channel spectrum allocation and reallocation. This is explained in next following section.

### 3.2. Proposed solution model and assumptions

We consider PUs as legacy device that accesses a block of spectrum through assigned static channels. This is a common scheme in various licensed spectrum bands. For example, a TV channel has a bandwidth of 6 MHz and there are more than 100 TV channels [11]. At any instant, each channel can be in either a busy or idle state, which refers respectively to times when a PU occupies the channel or does not. We assume the channel states are statistically independent and there are a total of $M$ frequency channels $\mathbf{F} = \{F_1,..., F_M\}$ in the CR network and there are total primary users $\mathbf{PU} = \{PU_1,..., PU_M\}$ in each of the frequency channels. These primary users can only occupy their assigned frequency channels. Since the primary users are licensed users, they are provided with an interference-free environment [4]. We assume that there are $N$ secondary users $\mathbf{SU} = \{SU_1,...,SU_N\}$ transmitting their own data in the system. Here **PU** and **SU** can be viewed as DNA sequences **Seq1** and **Seq2** respectively where as $\{PU_1,..., PU_M\}$ and $\{SU_1,..., SU_M\}$ are concerned nucleotides as described in preceding section. Average power of all the frequency channels $\{PU_1,...,PU_M\}$ and $\{SU_1,..., SU_M\}$ has to mapped with A=1, C=2, G=3 and T=4 of DNA sequence according to requirement of CR networks.

Let us consider for example, $M = 11$ and $N = 4$ i.e. there are total number of PUs and SUs are 11 and 4 respectively. The availability of channel for SUs depends upon idle state of PUs as shown in figure 5. In IEEE 802.22 a CR terminal is allowed to use a radio communication channel with a bandwidth of channel ranges 6 to 8 MHz in the frequency range between 41 MHz and 910 MHz [12].

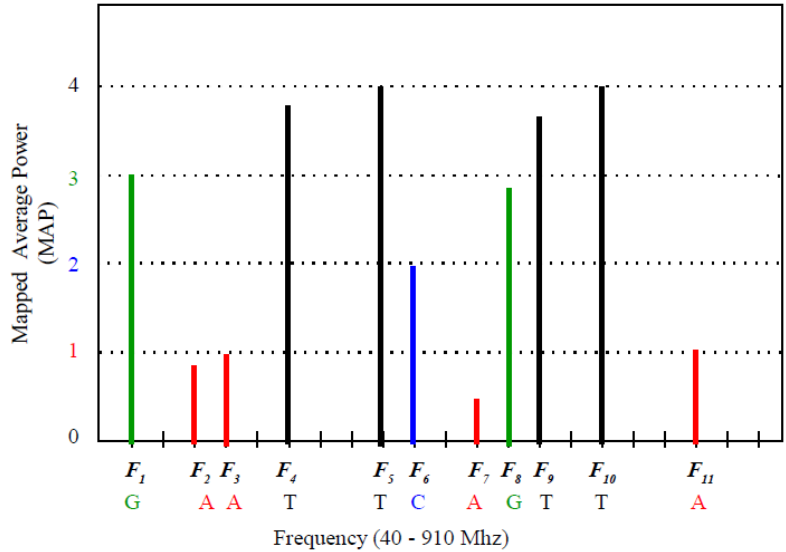

Figure 5

In figure 5, assumption has been made for channels of PUs as:

$F_i$ = A  if $0 \leq MAP \leq 1$ (Channels are idle or may contains noise); $F_i$ = C if $1 < MAP \leq 2$; $F_i$ = G if $2 < MAP \leq 3$; $F_i$ = T if $3 < MAP \leq 4$ ; where $i$ = 1, 2, … , $M$. So that sequence of PUs channel can be transformed to DNA sequence as $(Seq1)_{PU}$ = GAATTCAGTTA. Here nucleotide A is most required for allocating /reallocating channel spectrum of SUs.

On the SUs point of view, transmitter power of channels can be mapped with A, C, G, and T according to required power but should not exceed as direction of FCC [1], so that sequence of SUs channel can be formed as (Seq2)$_{SU}$. Therefore, these two sequence has been align using Needleman-Wunsch algorithm and can be adopted in physical [PHY] and media access control [MAC] layer of CR network to work efficiently for allocating/reallocating of SUs channel. Matlab simulation and results are demonstrated in next following section.

### 3.2. Simulation and results

In this section we have analyzed the simulation and its result based on NW optimization algorithm. Preceding examples of seq1= GAATTCAGTTA, and Seq2= GGATCGA of Trace-back matrix for alignment of these two sequences shown figure 6 and can be compared with figure 4.

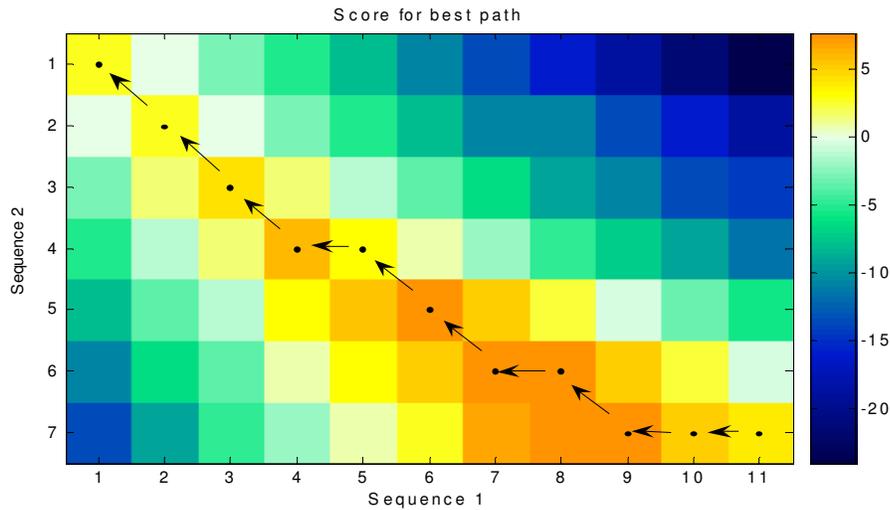

Figure 6

Let us consider at time instant $t_1$, PUs sequence assumed to be initially 11 available channels as Seq1= GAATTCAGTTA and simulation as shown in figure 7 can be compared with figure 5. It has been observed that there are 4 unused channels (ranges between unit 0 to 1) on PUs spectrum. SUs sequence assumed to be as Seq2 = AAAA for convenience and simulation of resultant alignment of Seq1 and Seq2 has been shown in figure 8. We observed that unused 4 channels of PUs are allocated to SUs channels shown as solid lines and broken line viewed as busy channels of PUs.

Now at time instant $t_2$, it is assumed that the PUs has changes dynamically as seq1= GAGTATCAGTA and simulation of resultant alignment of Seq1 and Seq2 (is assumed to be constant) as shown in figure 9. It can be easily observe that the availability of unused channels has been changed dynamically and reallocated to SUs with these unused channels.

Similarly, at time instant $t_3$, assumed that again PUs has changes dynamically as seq1= GAGTATCAATG and simulation of resultant alignment of Seq1 and Seq2 as shown in figure 10, which gives the same result as noted above.

We also examine a case where unused channels on PUs are less than the channels required by SUs. Let at any time instant Seq1= GAGTGTCAGTA and Seq2 = AAAA, resultant simulation is as shown in figure 11. We observed that only 3 channel for SUs has been allocated and remaining one channel is not allowed or has to be wait until PUs have enough unused channel.

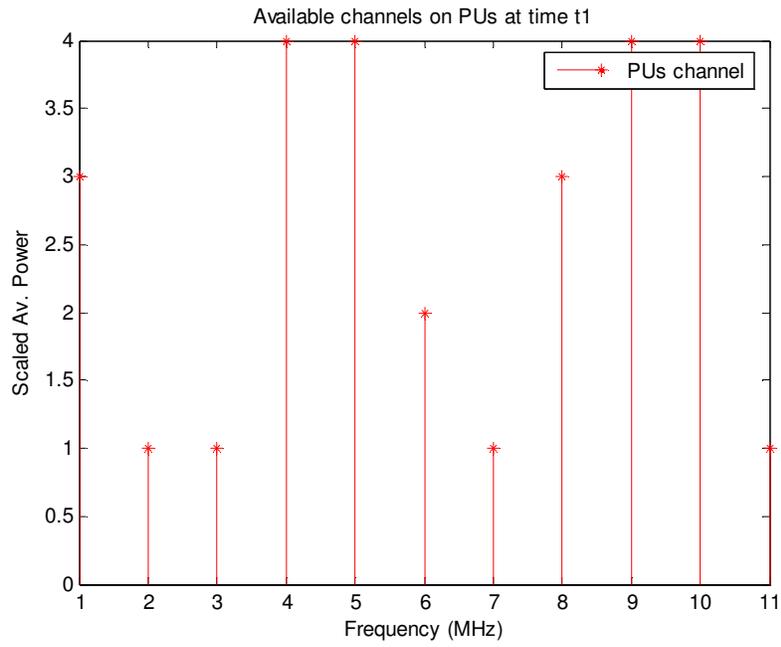

Figure 7

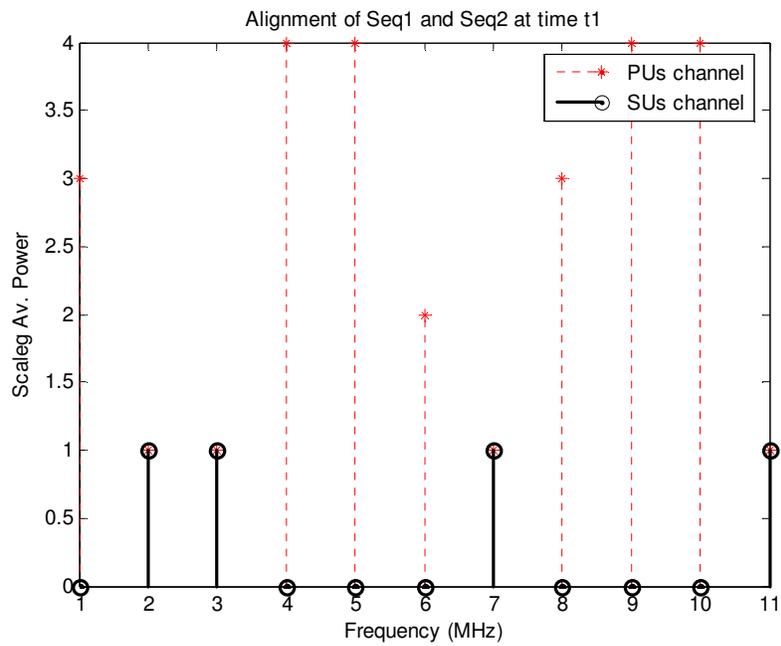

Figure 8

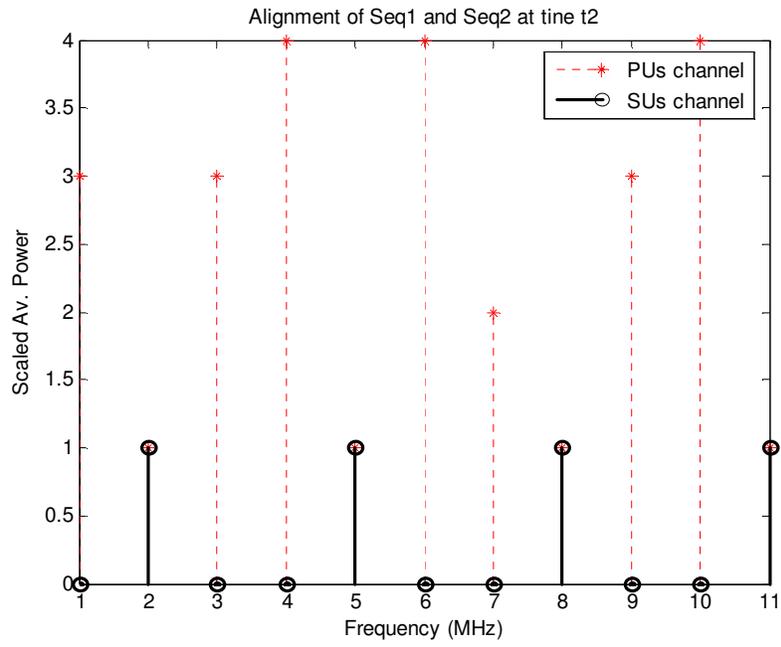

Figure 9

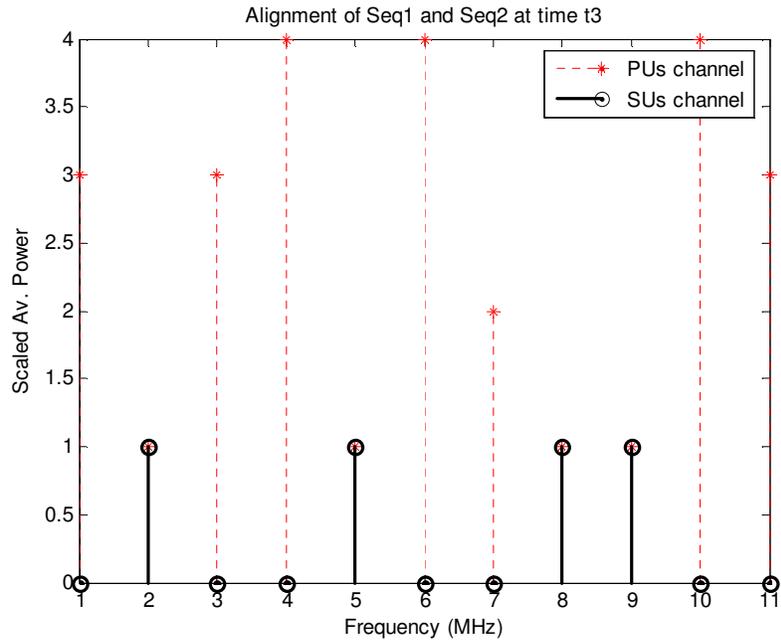

Figure 10

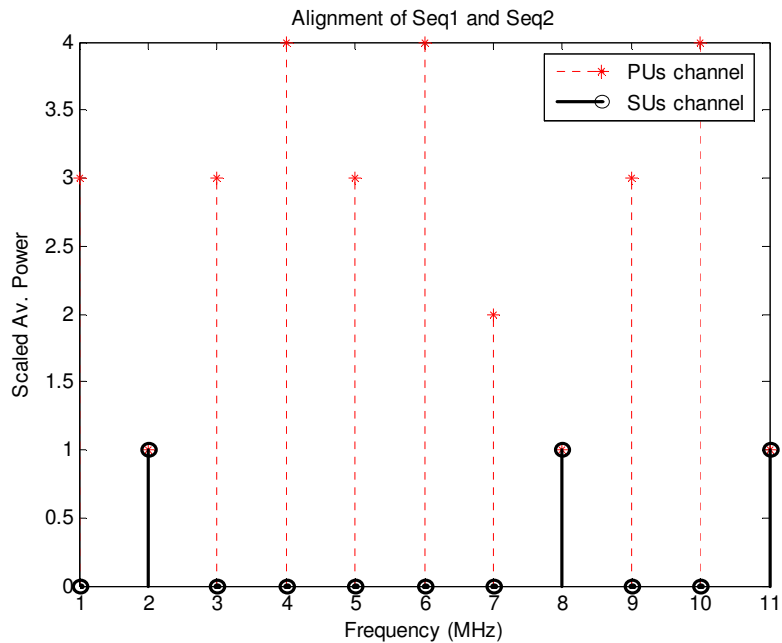

Figure 11

## 4. CONCLUSIONS

Researchers all over the globe are trying to find out the best solution to develop a wireless communications system that would be able to realize the requirements for a cognitive radio system. It has been seen that CR is the emerging spectrum sharing technology and can be the best option for next generation wireless networks since present spectrum crisis and uneven use of spectrum. In this paper, a DNA sequence alignment based opportunistic spectrum access scheme is proposed which is found to work fine. The best part is that spectrum allocation and reallocation condition in given Haykin's paper [4] as shown in figure 2 has been realize through figure 8, 9 and 10. The simulation programs for the proposed system are neither complex nor consume much time to respond. Hence, it can be easily embedded into physical [PHY] and media access control [MAC] layer of CR network to work efficiently. In this paper, however, most of communication parameter has been ignored and this model can make more robust by the inclusion of that parameters are proposed for future research work.

## ACKNOWLEDGEMENTS

We would like to thank the management who constantly inspired to involved in research work and special thanks to dean research Dr. S.C. Dwibedi for cordial cooperation and motivation in our work.

**Santosh Kumar Singh** received his B.E. degree in Electronics and Communication Engineering from S. J. College of Engineering, under Mysore University, Karnataka, India, year 1995 and M.Tech in Information Technology in 2004. He having 13 year teaching experience and pursuing his Ph.D. degree in Engg. at the School of Engineering, Suresh Gyan Vihar University, Jaipur, India. He published one book and one paper in well-reputed publication. He is also presented several paper in International and National conference. His current research interests include next generation wireless networks, wireless sensor networks and industrial embedded system.

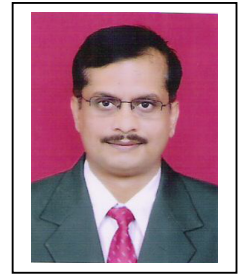

**Dr. Krishna Chandra Roy** received his M.Sc.(Engg) degree in Electronics and Communication Engineering from NIT Patna, Bihar, India and Ph.D degree in "Digital Signal Processing in a New Binary System" year 2003. He has currently Reader & Head of department (P.G.Engg.) of ECE, Sri Balaji College of Engg. & Tech., Jaipur, India and having 13 year teaching experience. He published two books Problems & solution in Electromagnetic Field Theory by Neelkanth Publishers (p) Ltd., Year-2006 and Digital Communication by University Science Press, Year-2009 respectively. He is also published and presented several paper in International and National conference. His current research interests include Digital Signal Processing and embedded system.

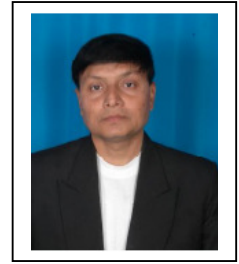

**Vibhakar Pathak** received his Master of Computer Application from Indira Gandhi Open University, India. He having 12 year teaching experience and pursuing his Ph.D. degree in Engg. at the School of Engineering, Suresh Gyan Vihar University, Jaipur, India. He is also presented several paper in International and National conference. His current research interests include next generation mobile wireless communication.

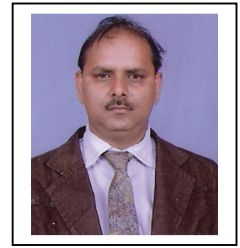